# Dynamical signatures of structural connectivity damage to a model of the brain posed at criticality

Ariel Haimovici and Pablo Balenzuela

*Departamento de Física, Facultad de Cs. Exactas y Naturales,*
*Universidad de Buenos Aires, Av. Cantilo s/n, Pabellón 1,*
*Ciudad Universitaria, 1428, Buenos Aires, Argentina and*
*Instituto de Física de Buenos Aires (IFIBA),*
*CONICET, Av Cantilo s/n, Pabellón 1,*
*Ciudad Universitaria, 1428, Buenos Aires, Argentina.*

Enzo Tagliazucchi[*]

*Netherlands Institute for Neuroscience, Meibergdreef 47,*
*1105 BA Amsterdam-Zuidoost, The Netherlands.*

*\* Corresponding author*
email: tagliazucchi.enzo@googlemail.com

**Running head:** Brain dynamics and structural damage.

**ABSTRACT**

Synchronization of brain activity fluctuations is believed to represent communication between spatially distant neural processes. These inter-areal functional interactions develop in the background of a complex network of axonal connections linking cortical and sub-cortical neurons, termed the human "structural connectome". Theoretical considerations and experimental evidence support the view that the human brain can be modeled as a system operating at a critical point between ordered (sub-critical) and disordered (super-critical) phases. Here, we explore the hypothesis that pathologies resulting from brain injury of different etiology are related to the model of a critical brain. For this purpose, we investigate how damage to the integrity of the structural connectome impacts on the signatures of critical dynamics. Adopting a hybrid modeling approach combining an empirical weighted network of human structural connections with a conceptual model of critical dynamics, we show that lesions located at highly transited connections progressively displace the model towards the sub-critical regime. The topological properties of the nodes and links are of less importance when considered independently of their weight in the network. We observe that damage to midline hubs such as the middle and posterior cingulate cortex is most crucial for the disruption of criticality in the model. However, a similar effect can be achieved by targeting less transited nodes and links whose connection weights add up to an equivalent amount. This implies that brain pathology does not necessarily arise due to insult targeted at well-connected areas and that inter- subject variability could obscure lesions located at non-hub regions. Finally, we discuss the predictions of our model in the context of clinical studies of traumatic brain injury and neurodegenerative disorders.

## I. INTRODUCTION

The human brain is never at rest, but constantly presents complex spatio-temporal patterns of coordinated activity instead (Raichle, 2006). The flexible and adaptive character of brain dynamics is required to cope with the environmental challenges of survival and reproduction, and most likely arises from evolutionary pressure (Cocchi et al, 2013). For instance, conscious perception of sensory information requires the brain to produce a widespread and temporally sustained response from the activation of a small number of cells in primary sensory cortices (Dehaene and Naccache, 2001). This extreme sensitivity to small perturbations can only arise if brain dynamics are intrinsically unstable. At the same time, however, brain dynamics cannot be characterized as fully random, since ordered patterns emerge in the way cortical regions interact with each other to support human cognition (Sporns, 2011).

This balance between unpredictability, instability and organized information processing can be modeled as a physical system posed at a critical point (Chialvo, 2010). The theory of phase transitions distinguishes different types of dynamics in physical systems undergoing a second order phase transition: super-critical dynamics are disordered, random and unpredictable, whereas sub-critical dynamics are ordered, regular and predictable. At the transition between both phases, a critical point exists at which dynamics present characteristics of both super-critical and sub-critical regimes (Chialvo, 2010). These features include an optimal exploration of transiently stable (metastable) states (Werner, 2007), the emergence of long-range coordinated activity fluctuations (Fraiman and Chialvo, 2012; Haimovici et al, 2013), a maximal sensitivity to external perturbations (maximal susceptibility) (Tagliazucchi et al, 2015a), and self-similarity which is manifest in the scale-free distribution of activity bursts (scale-free avalanches) (Beggs and Plenz, 2003). The critical point naturally accounts for synchronized activity without need of a hard-wired excitation/inhibition balance for the prevention of a massively over-sychronized state (Haimovici et al, 2013).

These properties are advantageous to the human brain and have been hypothesized to underlie its flexible and adaptive information processing capabilities (Chialvo, 2010). For instance, the maximal susceptibility at the critical point can account for the example provided in the first paragraph, i.e. a widespread response to a relatively small sensory perturbation. Brain states characterized by over-activation and over-synchronization (such as epileptic seizures) (Engel et al, 1982; Bartolomei et al, 2014) or by diminished activity/glucose metabolism (such as coma or the vegetative state) (Laureys et al, 2004) are pathological and depart from healthy conscious wakefulness, a state endowed with excitation/inhibition and correlation/anti-correlation according to the model of critical brain dynamics (Chialvo, 2010). It is natural to postulate that if the brain can be modeled as a critical system during health, injury and disease could be modeled as a displacement from such critical point.

While experimental evidence is compatible with the model of critical brain dynamics (Chialvo 2010, Tagliazucchi and Chialvo 2011), relatively few studies have addressed the hypothesis that brain injury can be modeled as a displacement from the critical point. This highlights the need of a modeling effort able to provide testable predictions for future experiments, as well as to explain available data in the light of this hypothesis. In the present work we filled this gap by constructing a conceptual large-scale model of critical brain dynamics unfolding on realistic anatomical or structural connectivity, and by using it to predict how several metrics related to different aspects of critical dynamics behave when the aforementioned network of structural connections ("connectome") (Hagmann et al, 2008) is damaged.

Previous computational modeling studies have attempted to link damage at the level of structural connections with functional impairments (Honey and Sporns, 2008; Alstott et al, 2009; Vása et al, 2015; Kraft et al, 2012). However, such studies have not investigated the connection between features predicted by critical models of the brain and structural damage, nor have they evaluated the consequences of lesions by focusing on brain-wide signatures of "healthy" flexible and adaptive dynamics (as opposed to investigating local changes in functional connectivity or activity). By virtue of the universality observed in critical dynamics (Chialvo, 2010) it could be possible to obtain general results independent of the microscopic details of previously investigated models. Also, while most previous studies focused on the functional alterations elicited by attacks targeted at nodes with special topological properties (for instance, *hubs*, i.e. nodes of high degree), it is currently unknown if similar alterations can be produced by removing a number of links equivalent to those lost when the hubs are removed. We investigate here this issue, which is crucial to evaluate the recent hypothesis that brain disorders preferentially target structural hubs (Crossley et al, 2014), and also for the interpretation of experimental reports linking lesions located at hubs with a variety of brain disorders (Buckner et al, 2009; Achard et al, 2012; Agosta et al, 2013; Rubinov and Bullmore, 2013; Baggio et al, 2014; Liu et al, 2014).

**II. MATERIALS AND METHODS**

  **A. Computational Model**

The computational model was adapted from previous work (Haimovici et al, 2013) and consists of a variation of the Greenberg-Hastings cellular automaton (Greenberg and Hastings, 1978). Connections in the model are given by the anatomical connectivity matrix (connectome). For a detailed description of the structural connectome and its derivation from diffusion spectrum imaging data see the work of Hagmann et al, 2008. Briefly, each of the 998 network nodes corresponds to a cortical region of approximately 2 cm$^2$ and the weighted links represent the density of white matter fiber tracts joining each pair of regions (measured with diffusion spectrum imaging (Hagmann et al, 2008).

Figure 1A shows the connectome embedded in anatomical space (left) as well as its weighted adjacency matrix (right), noted here by $W_{ij}$. The dynamical rules of the model are illustrated in Figure 1B. Each node may be in one of three possible states: inactve (I), active (A) or in a refractory state (R). The sequence of states followed by each node is dictated by three transition rules: (1) I → A with a small fixed probability $r_1 = 10^{-3}$, or if the sum of the connection weights with the active first neighbours is higher than a threshold T (i.e. node *i* becomes active if $\sum_{j,active} W_{ij} > T$), otherwise I→I (2) A→R always, and (3) R→I with a small probability $r_2 = 0.2$, delaying the transition from the R to the I state for some time steps. Parameters $r_1$ and $r_2$ were held fixed and determined the time scales of self-excitation and recovery from the excited state, respectively. Varying T allowed the system to follow a transition between a super-critical phase consisting of sustained but non-coherent activity (i.e. small T facilitated propagation of activity) and a sub-critical phase alternating periods of activation and quiescence, where spontaneous excitation could not become self-sustained (i.e. large T hindered propagation of activity). Figure 1C shows examples of the three possible scenarios (super-critical, critical and sub-critical). Examples of the mean activity produced by the model are shown in Figure 1D. The super-critical state is characterized by fast and temporally uncorrelated fluctuations, presenting a flat power spectrum typical of white noise (Figure 1D). The critical state presents oscillatory activity (the time scale of the oscillations is determined by $r_1$ and $r_2$) with long-range temporal correlations in their envelope. In this case the power spectrum has a well-defined peak and a slowly decaying tail. The sub-critical state also presents oscillations, but at a slower frequency and much more regular in terms of amplitude than in the critical case. Most of the spectral power is concentrated in the low frequencies. Previous work shows that this model best fits empirical fMRI data at its critical point (Haimovici et al, 2013).

**B. Metrics of model dynamics**

We explored a set of different metrics to describe the dynamics of the model and their distance to its critical point. These metrics were employed to assess the dynamical consequences of lesions on the structural connectome. Their choice was motivated by two considerations: i) they have been previously shown to behave differently in the super-critical, sub-critical and critical cases (either theoretically or empirically), ii) they have a heuristic interpretation in terms of brain activity and whether it can be considered healthy or pathological.

The following metrics were investigated:

- Lifetime of the sustained activity (τ) with spontaneous excitation suppressed ($r_1$=0). Simulations started with 10 random nodes in the active state, and the rest of the system inactive. The lifetime was computed as the number of simulation steps after which the activity stopped. This is the classical order parameter for the Greenberg-Hastings model (Greenberg and Hastings, 1978; Copelli and Campos, 2007).
- The mean and standard deviation of activity fluctuations:

$$<A>(t) = \frac{1}{N}\sum_{i=1}^{N} \delta(s_i(t),1) \quad (1)$$

$$<A> = \frac{1}{M}\sum_{t=1}^{M} <A>(t) \quad (2)$$

$$\sigma(A) = \sqrt{\frac{1}{M}\sum_{t=1}^{M}(<A>(t) - <A>)^2} \quad (3)$$

where N=998 is the number of nodes, M=300 is the length of the simulations and $s_i(t)$ the state of node $i$ at time $t$ with three possibilities: $s_i(t) = 0, 1, -1$ if the node is inactive, active and refractory, respectively. These metrics capture the degree of activity and whether global brain dynamics present temporal variability or are relatively constant over time Empirically, departures from a sufficiently high degree of variability have been linked to brain pathology and ageing (Garrett et al 2013).

- The size of the second largest cluster ($S_2$). Clusters consisted of neighbouring and simultaneously active nodes. "Neighborhood" was defined by the presence of a direct connection in the structural connectivity network. In contrast to the first largest cluster $S_1$ (which grows monotonously as activity increases in the model), $S_2$ peaks right before activity coalesces into a giant connected component, and decreases afterwards. These two metrics are standard order parameters in percolation theory (Stauffer and Aharony, 1994) and have also been employed as order parameters for the Greenberg-Hastings model (Haimovici et al, 2013). Biologically, $S_2$ captures how close the brain is to experimenting a massive co-activation, similar to a neural avalanche of activity (Beggs and Plenz, 2003; Tagliazucchi et al, 2012).

- The statistics of the pairwise correlation function. Each node's time series was binarized by assigning 1 when the node was active and 0 when it was either inactive or refractory. For purposes of temporal integration, we convolved the time series of each node with a standard hemodynamic response function (HRF) (Buxton et al, 2004) yielding for node $i$ a continuous signal $s_i(t)$ with mean value $\mu_i$ and standard deviation $\sigma_i$. The pairwise correlation matrix was computed as:

$$C_{ij} = \frac{<(s_i(t)-\mu_i)(s_j(t)-\mu_j)>}{\sigma_i \sigma_j} \tag{4}$$

and its mean and standard deviation as:

$$<C> = \frac{2}{N(N-1)} \sum_{i,j \neq i} C_{ij} \tag{5}$$

$$\sigma(C) = \sqrt{\frac{2}{N(N-1)} \sum_{i,j \neq i} (C_{ij} - <C>)^2} \tag{6}$$

These metrics capture global functional connectivity in the model ($<C>$) as well as the variability ($\sigma(C)$) (Deco et al, 2014a) of functional connections, i.e. whether all nodes have high or low coupling, or both strongly and loosely coupled nodes co-exist at the same time. Note that here the term "functional connectivity" is always used to denote statistical covariance between model time series and not between empirical fMRI time series.

- The linear correlation between the structural and functional connectivity matrices. We computed this in two ways. First, using both the weighted versions of the structural and functional connectivity matrices (C(F,S)$_w$). Second, correlating the binary (thresholded) versions of the matrices (C(F,S)$_b$). The structural connectome was binarized by assigning a value of 1 to all non-zero entries. The functional connectivity matrix was binarized by assigning a value of 1 to all entries larger than a certain threshold selected to fix the link density of the network. This resulted in binary versions of the structural connectome (Adj$_{ij}$) and functional connectivity matrices (B$_{ij}$). We used values for the threshold such that the link density ($\delta$) was between 0.01 and 0.3 (at steps of 0.01). The correlation between the binary adjacency matrices was averaged over the range of link densities.

$$C(F,S)_w = \frac{2}{N(N-1)} \frac{1}{\sigma(W)\sigma(C)} \sum_{i \neq j} (W_{ij} - <W>)(C_{ij} - <C>) \tag{7}$$

$$C(F,S)_b = <\frac{2}{N(N-1)}\frac{1}{\sigma(Adj)\sigma(B)}\sum_{i\neq j}(Adj_{ij}-<Adj>)(B_{ij}-<B>)>_\delta \tag{8}$$

It has been shown that an optimal exploration of the underlying connectivity takes place near the critical point of different computational models (Tagliazucchi et al 2015, Deco et al, 2014b). Empirically, changes in anatomy-function coupling can characterize the departure from normal conscious wakefulness (Tagliazucchi et al, 2015a, Tagliazucchi et al 2015b; Barttfeld et al, 2014).

- Susceptibility of the system ($\chi$). The system was perturbed by activating 60% of the nodes at time t=0 (i.e. synchronously transitioning 60% of the nodes -chosen at random- towards the active state, regardless of their previous state and the dynamics of the model)Each node's time series was convolved with the HRF, and the mean response of the system was obtained by averaging the simulated signals from all nodes. The variance of the mean response was computed in sliding windows of length equal to 20 time steps. The susceptibility was identified with the time elapsed until the variance decayed beyond a threshold of $10^{-3}$. Susceptibility is known to diverge for systems posed at criticality. The capacity of the brain to generate widespread and persistent activity in response to comparatively weak external perturbations (e.g. perceptual stimuli) has been postulated as a fundamental feature of conscious information access (Dehaene and Naccache, 2001). Experiments using non-invasive transcraneal magnetic stimulation (TMS) show that the elicited response is maximal during conscious wakefulness and is diminished in coma, vegetative and minimally conscious states (Casali et al, 2013).

- Inter-event time ($\Delta t_{ev}$), defined as the mean number of time steps between two consecutive activations. If a node was not activated during the whole simulation, its inter-event time was set equal to the length of the simulation (300 time steps). The inter-event time is a local measure of the frequency of activations and relevant for the simulation of structural lesions, since acute brain injury is known the slow down the frequency of activity recorded with electroencephalography (EEG) (Tebano et al 1988; Thatcher et al, 2001; Machado et al, 2004; Gaetz, 2004).

All results were averaged over 100 independent simulations with a total of M=300 time steps.

### C. Types of lesions

Each of the metrics defined above peaked at some intermediate point between the sub-critical and super-critical regimes. We studied how these transitions changed after the structural connectivity network lost connections due to targeted attacks. In particular, we explored which topological properties of nodes and links were most relevant to induce alterations in the dynamics after lesions. The topological properties studied here and their relevance for characterizing complex networks are discussed in Rubinov and Sporns, 2010.

We investigated the computational model unfolding over the structural connectome and progressively damaged by removing:

- Nodes with the largest degree (number of first neighbours).
- Nodes with largest connectivity strength (the sum of the weighted links to first neighbours).
- Nodes with largest betweenness centrality (Freeman, 1977). The betweenness centrality of node *i* is defined as the ratio between the number of shortest paths connecting two nodes passing through *i* ($\sigma_{jk}(i)$) over the total number of shortest paths ($\sigma_{jk}$). This was calculated using both the binary adjacency matrix and the weighted matrix. For the weighted case, we defined the distance between two nodes as the inverse of the weighted link connecting them.

$$BC(i) \frac{2}{(N-1)(N-2)} \sum_{j \neq k \neq i} \frac{\sigma_{jk}(i)}{\sigma_{jk}} \quad (9)$$

- Nodes with the largest eigenvector centrality. As an alternative measure of centrality, the eigenvector centrality of node *i* is equivalent to the i-th element in the eigenvector corresponding to the largest eigenvalue of the adjacency matrix (Lohmann et al, 2010).
- Nodes with the largest participation coefficient. The participation coefficient measures the importance of a given node for connecting different modules of the network (Guimera and Amaral, 2005).
- Links with the largest betweenness centrality.
- Links with the heaviest weights.
- Nodes chosen at random.
- Links chosen at random.

**III. RESULTS**

**A. Lesions at the 998 nodes resolution**

We first investigated the behavior of all metrics as a function of the threshold T, and how this behavior changed when nodes were targeted by degree (up to 300 of the nodes removed in order of highest degree, representing up to 65% of the connections in the network).

Figure 2 shows the results of this analysis for the metrics introduced in the Methods section ($\sigma(A)$: variance of total activity fluctuations, $\Delta t'_{ev}$: first derivative of the inter-event time, $\chi$: susceptibility, <C>: mean pairwise correlation function, $\sigma(C)$: standard deviation of the pairwise correlation function, $\tau'$: first derivative of the lifetime of the sustained activity, C(F,S)$_b$: correlation between anatomical and functional connectivity (binary), C(F,S)$_w$: correlation between anatomical and functional connectivity (weighted), S$_2$ : size of the second largest cluster). We observed that all metrics peaked at or near T=0.06, except $\Delta t'_{ev}$ that peaked at a higher value of T. This threshold value was very similar to the one previously reported as corresponding to the critical point of the model (Haimovici et al, 2013), even though the relatively small system size precluded the observation of a sharp phase transition. We note this critical threshold as T$_C$ = 0.06 and identified in our model with that of a "healthy brain", given that simulations are known to provide the best match with empirical fMRI data of healthy control subjects at T$_C$ (Haimovici et al, 2013). In all cases, removing nodes by largest degree displaced the curves to the left, i.e. all metrics progressively peaked at smaller values of T. This implies that for a fixed value of T, damage by degree will progressively drive the system out of the critical point towards the sub-critical regime.

We then studied how the critical threshold $T_C$ was displaced for all metrics as a function of targeted damage following the rules presented in the "Types lesions" subsection (quantified by the % of links removed). Results are shown in Figure 3. Targeted attacks should be compared to random removal of nodes and links. Interestingly, the only lesioning rules departing from random attacks corresponded to removing links by weight and by weighted betweenness centrality. Note that when removing links by weighted betweenness centrality, the damage rose up to only 20% of the network. This was because only 20% of the links had a non-zero weighted betweenness centrality (the distribution of weights was heavy tailed, so all shortest paths passed through a relatively small subset consisting of the "heaviest" links only).

Thus, the main results are that: i) all metrics except inter-event time consistently peaked at or near a critical threshold of $T_C = 0.06$, ii) the most relevant features to explain the dynamical consequences of structural damage were the weight and the weighted betweenness centrality of the removed links, iii) the effect of removing a large percentage of links by weight or by weighted betweenness centrality was to displace the system away from $T_C$ towards the sub-critical regime.

### B. Lesions at coarser anatomical regions

At the level of 998 nodes it is difficult to investigate the effects of coarse lesions encompassing larger brain regions with well-established functional roles. To investigate this, we mapped the 998 nodes onto the 90 cortical regions defined by the automated anatomical labelling (AAL) atlas (Tzourio-Mazoyer et al, 2002). Lesions in the AAL regions removed all connections with at least one end attached to any node within the region (containing a subset of the 998 nodes in the original parcellation). We note, however, that the simulations were still performed on the full network with 998 nodes and that the mapping onto the AAL atlas only determined the connections removed after lesioning the network.

To provide a single numerical index gauging the effect of lesioning the network, we first introduced the relative deviation in the value of each metric ($O$) with respect to its value at $T_C$ ("healthy" state, $O_0$):

$$\Delta O = \frac{O - O_0}{O_0} * 100 \qquad (10)$$

We ranked the regions according to the consequences of their removal by combining all metrics introduced in the "Metrics of model dynamics" subsection into a single damage index z, defined as:

$$z = \sqrt{\sum_O (\Delta O)^2} \qquad (11)$$

Figure 4A shows how each metric deviated from its value at $T_C$ after each region in the AAL atlas was targeted for removal. The red dots indicate the four regions with the largest damage score z. These regions correspond to the left and right precuneus (posterior cingulate cortex) and the middle cingulate cortex. The analysis performed in the previous section revealed that the most important features to explain changes in the metrics were the weight and the weighted betweenness centrality of the removed connections. To further explore this at the coarser level of anatomical regions in the AAL atlas, we studied the effects of randomly removing links with the constraint of preserving the sum of weights associated with the removal of each AAL region. In other words, we attempted to answer the following question: was there any topological property of the precuneus and middle cingulate cortex accounting for the dynamical consequences of removing them from the network, or was it possible to produce a similar damage by removing a larger number of links spread throughout the network, as long as their total associated weight equaled that of the aforementioned regions? The red curve in the bottom plot of Figure 4A shows that random attacks can cause an effect similar to targeted attacks, as long as the total weight of the removed links is the same.

We also investigated the similarity between results obtained using the different metrics. To this purpose, we computed the correlation coefficient between the absolute values of each pair of plots in Figure 4A. The absolute value was taken to discard the effects of the sign in the change of each measure. In fact, the z-index itself does not consider the sign of each measure either. A high correlation implies that the two metrics changed in a similar way when each AAL region was removed, a low correlation implies that the two metrics captured different aspects of the dynamical consequence of network damage. The results of this analysis are shown in Figure 4B in the form of a 12 x 12 matrix containing all correlation coefficients. The lower left corner of this matrix reveals that a number of metrics were highly correlated. The susceptibility $\chi$ was the most independent of all metrics, but $S_1$, mean activity ($<A>$) and the lifetime of sustained activity ($\tau$) were the most correlated with z-index, while $\sigma(A)$, $\sigma(C)$ and $<C>$ formed a 3 x 3 sub-matrix with high correlation values.

In Figure 5A we present a 3D rendering of the damage index z obtained by removing each AAL region. To investigate the relationship between z and the topology and weight of network connections at the coarser level of the AAL atlas, we considered the degree and betweenness centrality as well as their weighted counterparts (strength and weighted betweenness centrality). These network features were calculated on the coarser 90 x 90 adjacency matrix derived from the finer grained 998 x 998 original matrix. The *i,j* entry in this matrix was computed as the sum of all links with one end in any node within region *i* and the other end in any node within region *j*. A rendering of these features onto brain anatomy is presented in Figure 5B. As expected from the results discussed in the previous paragraph, all network features taking into account link weight presented a higher correlation with z (Figure 5C).

All analyses performed up to this point concerned global averages (i.e. how damage to a single region impacted the dynamics of the system as a whole). It is interesting to investigate if specific effects could be obtained by damaging each AAL region. To this aim we investigated the changes in inter-event time ($\Delta t_{ev}$) in response to removing each of the 90 AAL regions. The results of this analysis are shown in Figure 6A. As shown in this figure, the weight of the link connecting each pair of regions predicted $\Delta t_{ev}$ after removing one of them, especially for large ($>10^{-1}$) connection strength.

Figure 6C shows renderings of $\Delta t_{ev}$ in all 90 regions after removing each of the top ten regions in terms of the global change of $\Delta t_{ev}$ (as in Figure 4A). In all cases we observed that the local lesion affected a relatively widespread network of regions, and that the most affected regions were anatomically close to the location of the lesion. Changes in $\Delta t_{ev}$ were always in the direction of a slowing down of activity (i.e. increases in inter-event time). The renderings presented in Figures 5 and 6 were visualized with BrainNet Viewer (http://www.nitrc.org/projects/bnv/) (Xia et al, 2013).

## IV. DISCUSSION

Prior to the development of neuroimaging, the study of brain lesions was possibly the only way of systematically probing the functional role of cortical regions (Posner, 1988). The development of methods capable of resolving the large-scale connectivity of the human brain, together with theoretical and computational modeling advances, allows an exhaustive *in-silico* exploration of the consequences of anatomical lesions. While the study of behavior is not possible from simulations, they can be used to predict alterations in brain dynamics as a function of the topological properties of the damaged nodes. The uncontrolled nature of the lesions investigated in clinical studies make an exhaustive exploration of cortical lesions very difficult or even impossible. This exhaustive exploration is only possible through the simulation of focal structural damage in models of brain dynamics.

The present study investigated the dynamical consequences of introducing lesions (i.e. removal of network nodes) in a simple model of whole-brain activity posed at the critical point of a phase transition. In contrast with previous studies using more complex and realistic models of neural dynamics (Honey and Sporns, 2008; Alstott et al, 2009), our approach allowed us to focus on how lesions displaced the system from such critical point, and how this displacement impacted on a number of metrics of "healthy" brain dynamics. Our main conclusion is that attacks targeted at high-weight connections disrupted the dynamics of the system by decreasing its critical threshold $T_C$ and thus inducing a sub-critical state. The emergence of the sub-critical state after structural network damage implies a decrease in the activity levels of the system and a general slowing down of its fluctuations, loss of functional connectivity and its variability, and diminished reactivity to external perturbations (susceptibility). While experimental evidence suggests that the human brain can be modeled as a system at criticality, (Chialvo, 2010; Tagliazucchi and Chialvo, 2011), relatively few studies have modeled the effects of brain lesions and disorders from this perspective. Future research efforts should focus on evaluating signatures of sub-critical dynamics after human brain injury, as well as other signatures accessible from invasive neural recordings in animal models (e.g. alterations in the scale-free distribution of neural avalanches) (Beggs and Plenz, 2003; Tagliazucchi et al, 2012).

An ongoing controversy exists concerning the role that criticality plays in human brain function. In first place, the concept of criticality originates from physical systems with relatively simple and well-understood rules; it is thus unclear whether it applies to more complex biological systems (such as the brain) or whether it is only useful as a model or approximation. Furthermore, critical dynamics could be dismissed as an uninteresting consequence of self-organized non-linear systems with many degrees of freedom (Bak et al, 1987). However, an important role for criticality is suggested by studies showing that slow wave sleep (Priesemann et al, 2013), epilepsy (Meisel et al, 2012), and anesthesia (Scott et al, 2014) can be modeled as a departure towards sub- and super-critical regimes. Experimental confirmation of our prediction – modeling brain injury as a displacement towards the sub-critical state - would add further support to the importance of criticality. It has also been questioned whether the dynamics of the brain during wakeful rest can be considered critical (Beggs and Timme, 2012), and it has been suggested that slightly sub-critical dynamics could provide a "safety margin" necessary to avoid the pathological consequences of super-criticality (Priesemann et al, 2014). More generally, it has been argued that the widespread observation of criticality in complex systems could be biased by the inference methods (Mastromatteo and Marsili, 2011) and the selection of statistically relevant descriptions of the data (Marsili et al, 2013; Haimovici and Marsili, 2015). However, our study must be understood as a first approximation to the question: "*What are the dynamical consequences of structural damage to a model of brain dynamics posed at criticality?*". Future studies must address whether the brain is really critical, slightly sub-critical, or whether criticality in the brain is global or local, and whether the dynamical regime of the brain is fluctuating or in a steady state. It is important to discuss our results in the context of previous empirical and simulation studies. Modeling of focal damage to the human and macaque structural connectomes revealed a maximal impact of lesions located at midline and parietal regions, coinciding with the nodes of highest degree (hubs) (Honey and Sporns, 2008; Alstott et al, 2009; Cabral et al, 2012). This might follow from the observation that removing hubs from the connectome results in maximal structural disconnectivity, as measured with the small-world index (Sporns et al, 2007). Since hubs are the nodes associated with the highest number of connection fibers, our finding that the main predictor of the dynamical consequences of network damage is the number of affected fibers is consistent with these previous studies. While most previous work focuses on the relationship between global functional or structural network properties and the topological role of the lesioned nodes, the study of Váša and colleagues focuses on the consequences of damage to a simple conceptual model exhibiting a critical point (the Kuramoto model) (Váša et al, 2015). Their finding of increased metastability after damage to hubs apparently contradicts our observation of a shift towards the sub-critical regime. It is not clear, however, that their tuning method results in parameters corresponding to the critical point of the model (as measured, for instance, by the divergence of susceptibility); therefore a displacement towards sub-critical dynamics from a super-critical working point could increase the flexibility of the dynamics (metastability).

The changes in the dynamics after simulated damage were consistent with empirical studies of acute brain injury, for instance, of traumatic brain injury and acute ischemic stroke. Metabolism is known to decrease after these events, and decreased glucose consumption is indicative of diminished energetic demand and neural activity levels (Kuhl et al, 1980; Baron et al, 1986; Vagnozzi et al, 2010). A focal slowing down of EEG rhythms is generally observed in the affected region and might also extend to a larger cortical network, often including the region contralateral to the lesion site (Tebano et al 1988; Thatcher et al, 2001; Machado et al, 2004; Gaetz, 2004). Reactivity (e.g. reaction time) can be decreased after brain injury, suggesting a less efficient percolation of sensory information to the executive and motor networks of the brain (Stuss et al, 1989). Traumatic brain injury and stroke both reduce long-range functional communication in the brain (assessed with resting state fMRI) (Mayer et al, 2011; Sharp et al, 2011; Grefkes and Fink, 2014). These empirical observations are consistent with the drop in activity levels ($\tau$ and $<A>$), frequency of activations ($\Delta t_{ev}$), susceptibility ($\chi$) and pair-wise correlation ($<C>$), respectively, characterizing the displacement towards the sub-critical regime. It must be kept in mind that brain metabolism was not directly modeled and thus only an indirect link may exist between it and certain simulated variables, such as the rate of activations of a given node.

Brain dynamics and metabolism are fundamentally altered in the extreme of damage leading to a state of non-responsiveness, such as coma or the persistent vegetative state. In these cases, fronto-parietal metabolism (Laureys et al, 2004) and functional connectivity (Vanhaudenhuyse et al, 2010) are diminished, and the power of slow EEG rhythms (such as delta at 1-4 Hz) is globally increased (Schiff et al, 2014). A typical EEG pattern observed in comatose patients is *burst suppression*, consisting of periods of activity ("spikes") alternating with periods of inactivity (Young, 2000). This pattern is qualitatively similar to the global average produced by our model in the sub-critical regime (see Figure 1D, third panel), and consistent with our observation that extensive structural damage leads to sub-critical dynamics in the model. A drastic reduction in the susceptibility $\chi$ of the system could cause a failure of sensory stimuli to elicit the widespread brain activation postulated to underlie conscious access by the global workspace theory of Baars and Dehaene (Dehaene, 2001), accounting for loss of conscious content after severe brain injury. The bistable nature of sub-critical dynamics (periods of activity alternating with periods silence) presents a low degree of *differentiation*, understood as a highly reduced number of possible states. Reduced differentiation in the sub-critical regime is consistent with loss of conscious awareness in the context of the Information Integration Theory of consciousness put forward by Tononi (Tononi, 2004).

Another consequence of structural damage to our model operating at criticality was decreased similarity between functional and structural networks. This prediction is valuable since alterations in the coupling between anatomical and functional networks remain understudied in the context of brain disorders. Research on human slow wave sleep and anesthesia, two brain states modeled as sub-critical (Priesemann et al, 2013; Scott et al, 2014), reveals a regional decoupling of brain structure and function (Tagliazucchi et al, 2015a; Tagliazucchi et al, 2015b) supporting the prediction that brain lesions could result in a similar decoupling.

Clinical studies reveal that the effects of localized damage (e.g. stroke) are highly variable and depend on the location of the insult (Damasio and Damasio, 1989). Our analyses highlighted a small set of high degree regions (hubs) as the most vulnerable points of brain anatomy. In particular, targeting midline regions such as the posterior cingulate cortex/precuneus and the middle cingulate cortex maximized the dynamical consequences of the lesion. This observation resonates with a recent meta-analysis showing that brain disorders tend to be associated with anatomical damage (quantified via voxel-based morphometry) located at structural hubs (Crossley et al, 2014). These regions are generally considered "higher-level" cortical areas (as opposed to "lower-level" primary sensory areas) and operate as converging or integrative areas in the brain. In the case of the posterior cingulate cortex, it is also a pivotal node in the default mode network (DMN) (Fransson and Marrelec, 2008) - a network of brain regions increasing its activity during rest (Raichle et al, 2001) and implicated in a variety of brain disorders (Whitfield-Gabrieli and Ford, 2012). The high energetic cost of hub regions (Liang et al, 2013) could augment their vulnerability, as they would be the first to be affected by brain pathologies restricting neuronal metabolism.

One important question left unanswered by previous modeling studies is whether the removal of hubs is equivalent to the removal of a set of links of equivalent weight. Equivalently, do brain hubs have some special topological role in the connectome so that their removal has the strongest effects on dynamics, or can we obtain a similar result by lesioning a larger number of smaller-degree nodes? Here we demonstrate that for our model of brain dynamics at criticality the second alternative is correct (Figure 4A, bottom plot). This observation suggests a different reason why hubs are involved in such a wide spectrum of brain disorders. It is possible that certain brain pathologies (either acute or neurodegenerative) arise due to widespread damage amounting to a certain weight of structural connections. However, inter-individual variability will obscure differences that are not located at hubs, since many possible combinations exist to remove connectome links summing up to a certain total weight. On the other hand, damage to cortical hubs provides a consistent way of removing a large connectivity weight from the network, and therefore these lesions will be highlighted after averaging over large clinical populations. This hypothesis can also explain why the default mode network appears implicated with such a wide spectrum of dissimilar diseases (Whitfield-Gabrieli and Ford, 2012), since it is mainly comprised by high-degree regions (Hagmann et al, 2008). Since it is known that individual structural connections are not sufficient for the prediction of the emergent functional networks (Mišić et al, 2016), future studies should focus on evaluating the impact of removing structural connections on the formation of coordinated large-scale functional networks, and whether this impact also only depends on the number of removed fibers, regardless of their topological role.

Our study has a number of limitations to be addressed in the future. First, diffusion spectrum imaging is an imperfect technique for determining the structural connectivity of the human brain. In particular, it is known to under-estimate long-range and homotopic connections (Messé et al, 2014; Reveley et al, 2015). As the diffusion MRI techniques are improved over time, simulation studies based on the structural connectome should be revisited. A related limitation is the reduced number of subjects used to create this version of the structural connectome. While modifications in the underlying connectivity of the model could change how regions are ranked according to their vulnerability (Figure 4), it is unlikely that other aspects of our results will be affected; for instance, our result concerning the similarity between targeting hubs or an equivalent number of lower-degree nodes. A second limitation is the lack of plasticity in our model. The human brain is known to reorganize and compensate for anatomical damage (Kolb and Gibb, 2007), and while a displacement towards the sub-critical regime is predicted by our analysis, it is likely that changes in connectivity over time will revert the dynamics of the brain towards the critical point. It is also likely that brain injury leads to the release of neurotransmitters capable of globally altering brain excitation (e.g. glutamate) (Katayama et al, 1990), represented in our model as a shift of the threshold towards a critical value. Finally, the displacement towards the sub-critical regime after network lesions could be predicted from the absence of inhibitory connections, whose disruption could lead towards super-critical dynamics instead. However, diffusion spectrum imaging and related techniques cannot differentiate between long-range excitatory and inhibitory connections (Park and Friston, 2013). Furthermore, the majority of inhibitory connections in the brain are local and do not project over distant cortical areas (Freund and Kali, 2009). In our model, inhibition of local nodal dynamics was represented by the probability of transitioning out of the refractory period, a parameter that was held constant in all simulations.

Finally, we employed a simple and non-realistic model of brain activity. However, the universality of critical dynamics could guarantee that similar results will be observed for more complex and detailed computational models of neural dynamics, as long as they are posed at a critical point. In this sense, our simple model is an advantage: studying the effects of anatomical lesions by means of all known variations of neural models would represent a painstaking effort; however, we were able to provide generic results by focusing on the concept of criticality instead of addressing the microscopic details of the model.

**V. CONCLUSION**

In summary, we have combined empirical information on the network of human anatomical connections with the hypothesis that brain dynamics can be modeled as a critical point of a phase transition. From this combination we derived a number of results consistent with existing studies, and made clear predictions for future experiments. The hypothesis of criticality as a plausible model for brain dynamics should not only attempt to explain the defining characteristics the healthy state, but also account for the consequences of anatomical damage. Our work represents a first step in this direction and should be followed by experiments designed to test its predictions, both in human subjects and animal models of brain pathology.

ACKNOWLEDGMENTS

We thank Olaf Sporns and Patric Hagmann for sharing the structural connectivity network and the Brain Connectivity Toolbox (BCT) (available at https://sites.google.com/site/bctnet/). A.H. is supported by a CONICET doctoral fellowship. E.T. is supported by an AXA post-doctoral fellowship.

AUTHORS DISCLOSURE STATEMENT

No competing financial interests exist.

**FIGURES**

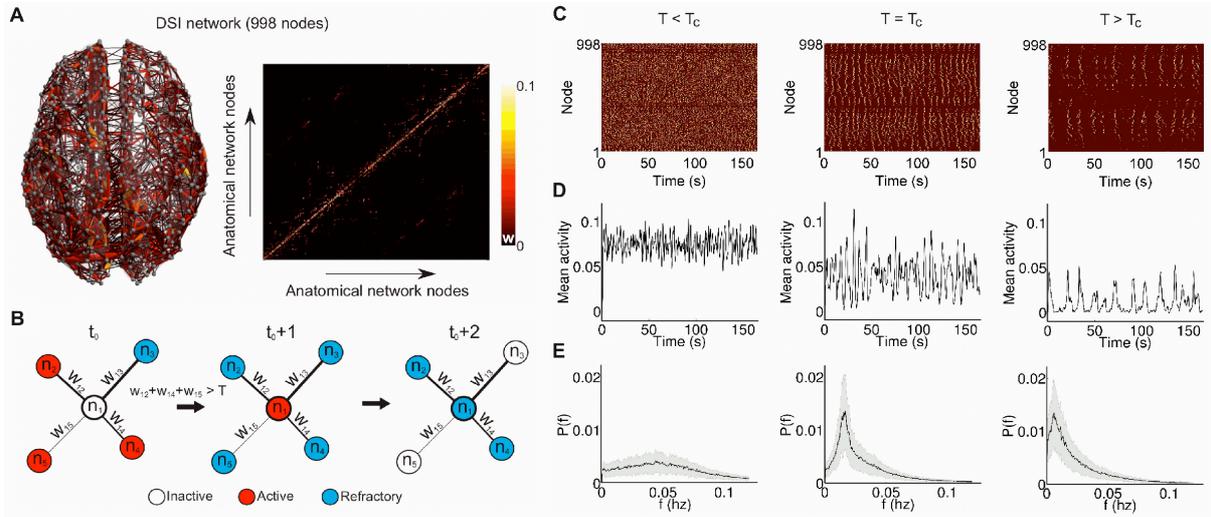

**Figure 1**: Computational model and the underlying connectivity network. A) Anatomical embedding of the structural connectome measured with diffusion spectrum imaging (left) and its adjacency matrix $W_{ij}$ (right). B) Illustration of the rules governing the transition between the three possible states of the system: inactive (I), active (A) and refractory (R). In the example, three active nodes and one refractory node surround an inactive node; activity propagates to the central node in the second time step, and finally this node reaches the refractory state in the third step. C) Example of super-critical ($T < T_c$), critical ($T = T_c$) and sub-critical ($T > T_c$) dynamics for all 998 nodes simulated for 150 time steps. D) Examples of the average activity generated by the model in the super-critical, critical and sub-critical regimes. E) Power spectra (mean ± SD) of the average activity generated by the model in the super-critical, critical and sub-critical regimes.

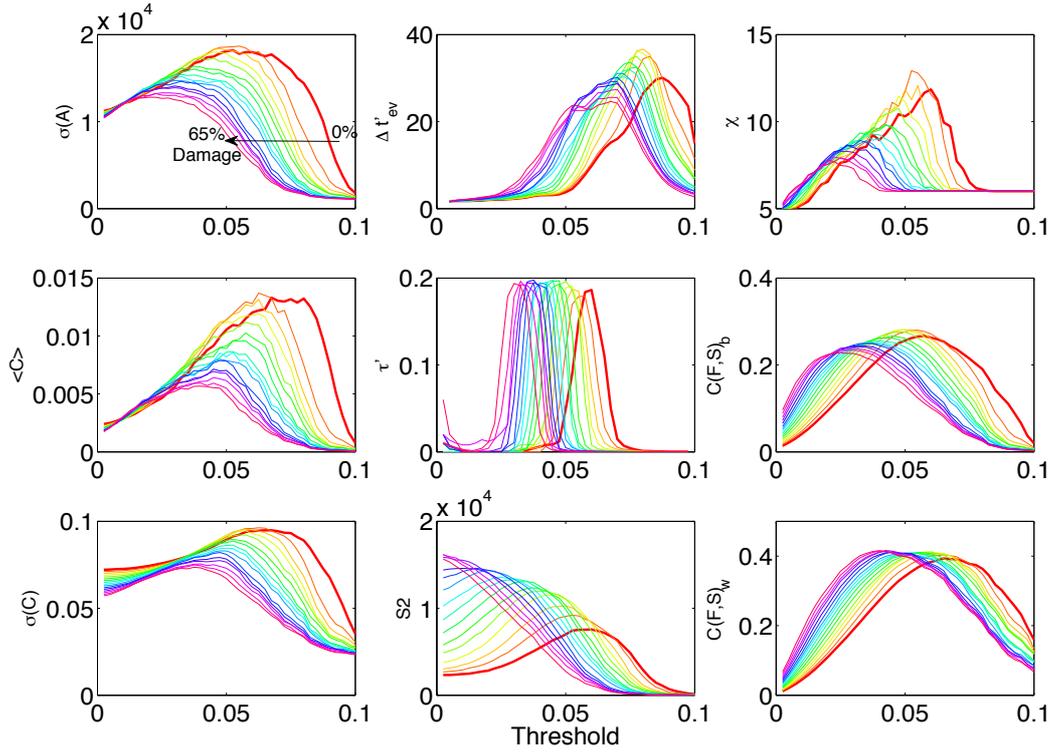

**Figure 2**: Changes in the phase transition as a response to attacks targeting the highest degree nodes. We disconnected the nodes with largest degree, up to 300 nodes (equivalently, 65% of the connections in the network). For each metric, the critical point moved from its original position (thick red curve) to lower values. The inter-event time ($\Delta t_{ev}$) presented a monotonic increase saturating at large $T$ and vanishing for small $T$. Similarly, the lifetime of the activity ($\tau$) saturated for small $T$ values and vanished for large $T$. We therefore plotted the derivative of these metrics with respect to $T$ ($\Delta t'_{ev}, \tau'$) to obtain their critical values.

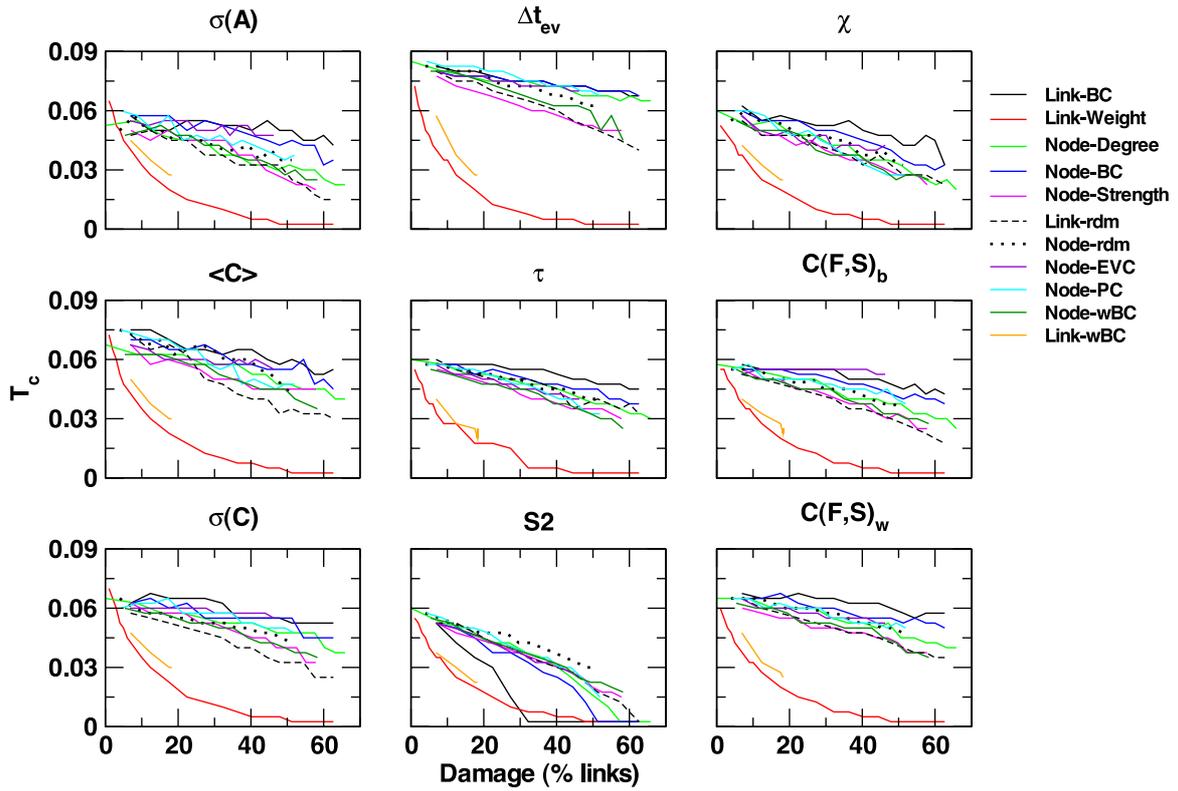

**Figure 3**: Displacement of the critical point as a result of damage following different procedures. The critical point ($T_C$) moved towards lower values when the network was damaged following all the explored criteria. We removed links and nodes by largest binary (BC) and weighted (wBC) betweenness centrality, largest degree, connectivity strength, eigenvector centrality (EVC) and participation coefficient (PC), as well as links by largest weight. In addition, the dotted and dashed lines show the results of randomly removing nodes and links. The only two criteria leading to damage significantly greater than random attacks were the removal of links by weight and by weighted betweenness centrality.

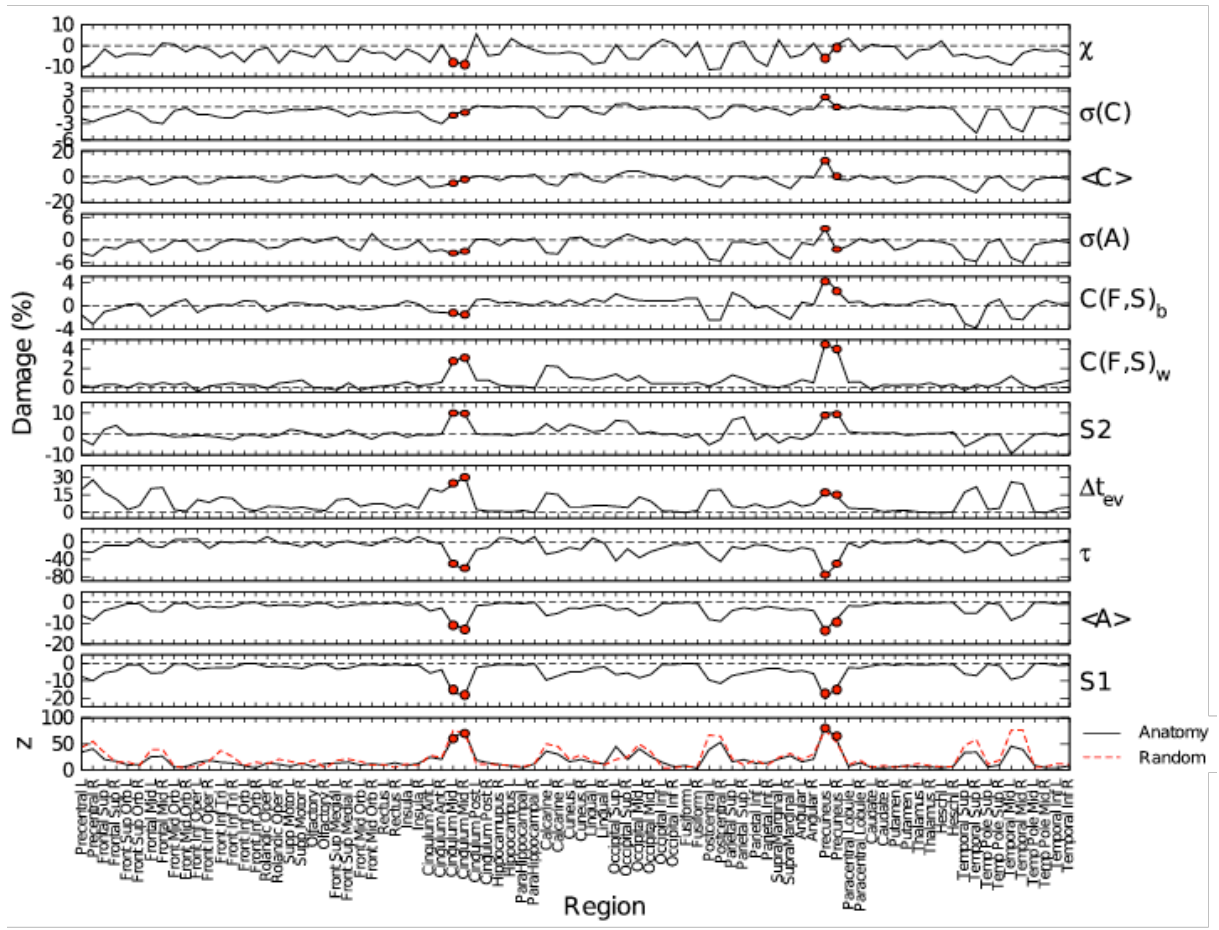

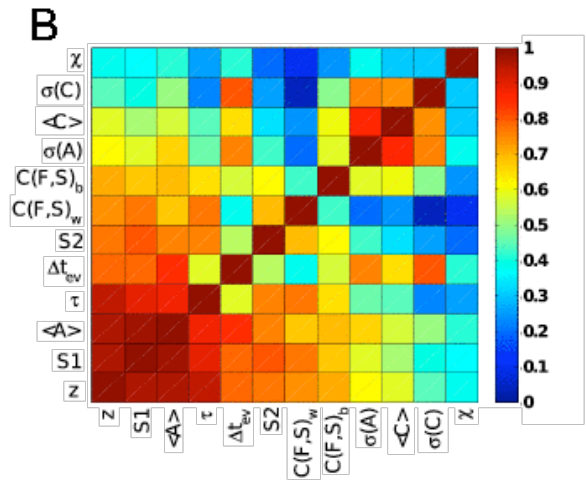

**Figure 4**: Dynamical consequences of lesions at the AAL atlas resolution. A) Response of the system after attacking each of the 90 AAL regions. Each sub-plot shows the change in each metric relative to the un-attacked system (dashed line at level zero). The bottom plot shows the damage score z (Eq. 11) for each region. Red dots highlight the regions presenting the largest z (bilateral precuneus and middle cingulate cortex). The red curve in the bottom plot shows the results obtained by randomly removing links with the constraint of having a total weight equivalent to that of the corresponding AAL region. B:) Correlation matrix for the changes in the metrics across all AAL regions.

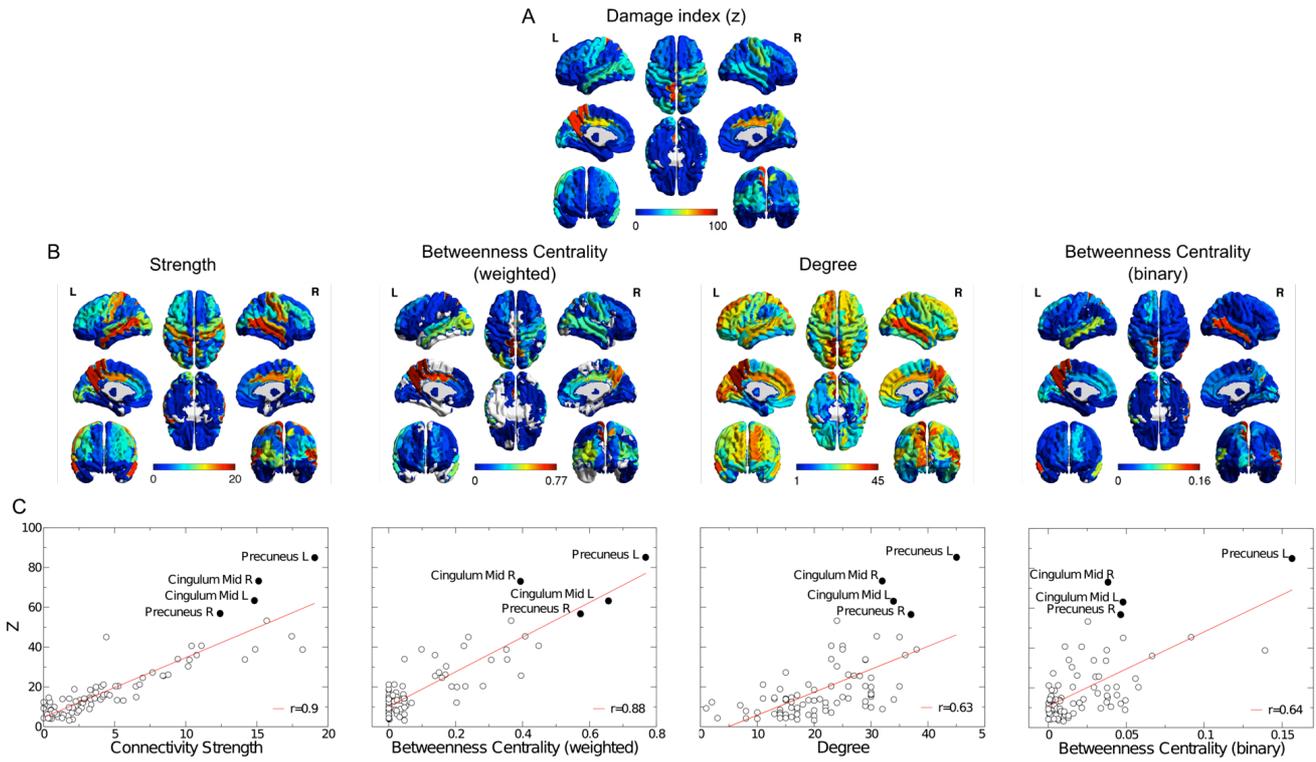

**Figure 5**: Correlation between the damage index z and the topological features of the network. The top rendering (panel A) shows the values of z across all 90 AAL regions. The bottom renderings (panel B) show the values of strength, weighted betweenness centrality, degree and binary betweenness centrality. C) Scatter plots and linear fits for the damage index z vs. the topological properties illustrated in panel B. Weighted measures (strength and weighted betweenness centrality) show the highest correlations with z.

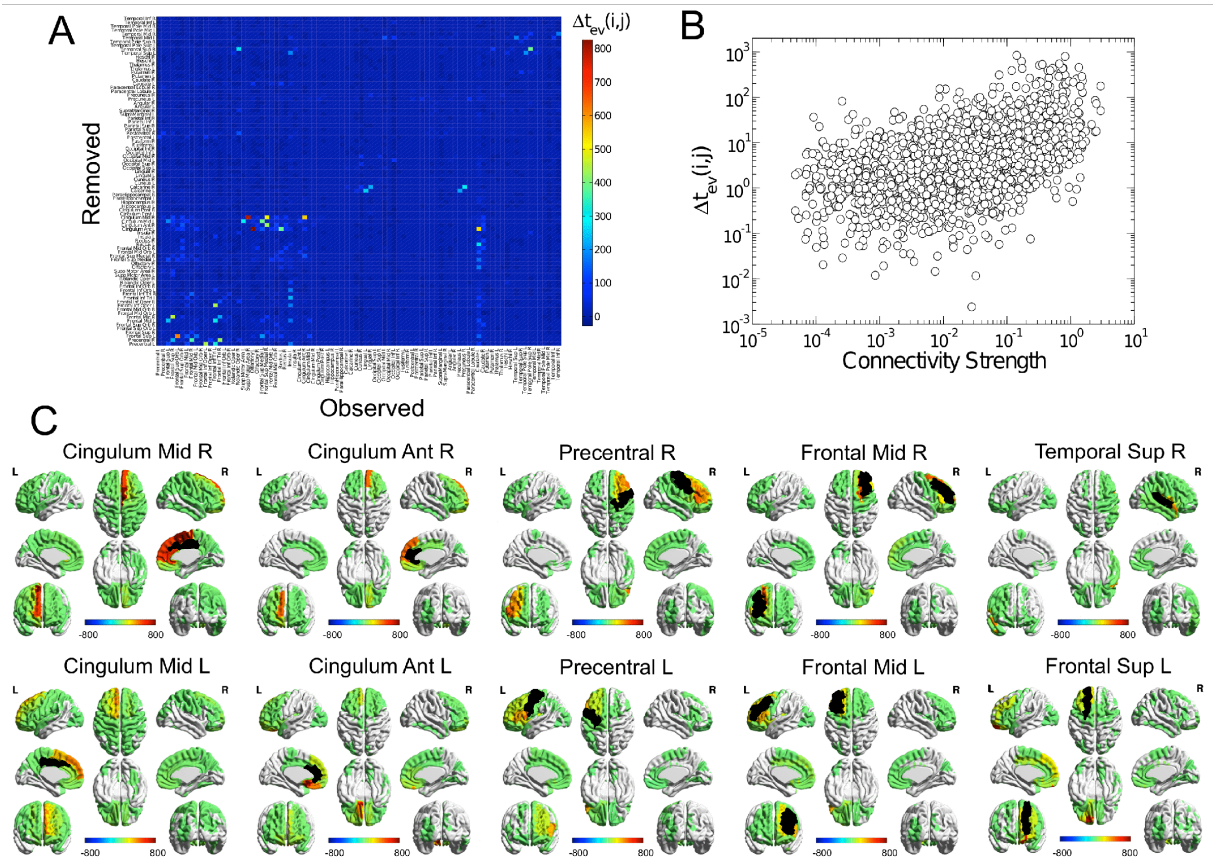

**Figure 6**: Consequences of lesions on the local dynamics of AAL regions. A) Local changes in the inter-event time ($\Delta t_{ev}$) at each AAL region (columns, "Observed") in response to removing another AAL region (rows, "Removed"). B) Changes in inter-event time as a function of the strength of the link joining each pair of regions. C) Renderings of the changes in inter-event time after removing the top 10 regions in terms of their influence after targeted lesions.